\documentclass[a4paper,12pt]{article}
\usepackage{graphicx}				
\usepackage{subcaption} 
\usepackage{comment}
\usepackage[backend=bibtex,style=ieee,]{biblatex}
\usepackage{caption}
\usepackage{amssymb}
\usepackage{amsmath}
\usepackage{amsthm}
\usepackage{bm}
\usepackage{color}
\usepackage{authblk}
\usepackage{mathrsfs}

\newcommand{\pow}{n}
\newcommand{\B}[1]{B_{#1}}
\newcommand{\x}[1]{\rho_{#1}}
\newcommand{\ph}[1]{\phi_{#1}}
\renewcommand{\th}[1]{\theta_{#1}}
\title{Spectral Analysis of Scattering Resonances with Application on High Contrast Nanospheres}
\date{}
\newtheorem{lem:convergence}{Lemma}[section]
\newtheorem{thm:eigenvalueAppr}{Theorem}[section]
\newtheorem{thm:eigenvalueAppr1}{Theorem}[section]
\newtheorem{thm:eigenvalueAppr2}{Theorem}[section]
\newtheorem{cor:spcase}{Corollary}[section]
\newtheorem{cor:shari}{Corollary}[section]
\newtheorem{cor:nano}{Corollary}[section]
\author[1]{Brian Adams}
\author[1]{Kevin Li}
\author[1]{Taoufik Meklachi}
\affil[1]{School of Science, Engineering, and Technology, Penn State Harrisburg}
\medskip
\begin{document}

\maketitle

\begin{abstract}
In this paper we provide further spectral analysis of the general asymptotic scattering resonances formula of small 3D dielectrics of arbitrary shape with high contrast, initially derived to a first order approximation in \cite{Meklachi2018}. To investigate the components of a full expansion of such resonances, a breakdown is presented for the case of high contrast nanospheres. We also derive, for radially symmetric fields, an exact resonance formula for a spherical scatterer in terms of its radius, not necessarily small, and dielectric susceptibility coefficient, not necessarily high. Such a formula is useful in imaging applications to identify objects' properties from frequency measurements. This formula is further developed and simplified in the case of high contrast nanospheres.    
\end{abstract}

\section{Introduction}
Spectral analysis of scattering resonances has been the center of interest in many branches of medicine, physics, mathematics, and engineering. Scattering resonances arise in characterizing the rates of oscillation and decay of scattering waves, encoded, respectively, in the real and imaginary parts of the associated eigenvalues.

In this work we present spectral formulas, exact and asymptotic, of potential  scattering of three dimensional linear media in a vacuum. The governing equation is the Helmholtz equation $\Delta u+Vu = 0$, subject to Sommerfeld radiation condition at infinity, where   $V$ is compactly supported in $\mathbb{R}^3$ and is allowed to be complex valued, and $u$ represents the scalar fields. The general theory behind computing resonances via  analytic expansions and some related findings and applications can be accessed in~\cite{Zworski2016}\cite{Dyatlov2019}\cite{Colton1992}. This work is an expansion of the results in~\cite{Meklachi2018}, where non-linear eigenvalue problems corresponding to scattering resonances of small high contrast volumes in linear and non-linear media are solved asymptotically to the first order. One goal of this development is to contribute to the theory of wave manipulation at subwavelength scales. This artificial capability allows for the manufacturing of novel materials, such as metamaterials, designed at the nanoscale level. Due to the high resonances at the surface, such materials manifest superb phenomena, such as cloaking~\cite{Meklachi2016}\cite{Milton2006}\cite{Bouchitt2010}\cite{Bruno2007}\cite{Cai2007}\cite{Greenleaf2009}\cite{Vasquez2009}\cite{Kohn2010}\cite{Kohn2008}\cite{Lai2009}\cite{Liu2009}\cite{McPhedran2009}\cite{Miller2006}\cite{Milton2008}\cite{Schurig2006} and superlensing~\cite{Milton2005}\cite{Nicorovici1994}\cite{Al2005}\cite{Bryan2010}\cite{Pendry2000}\cite{Zhang2008}\cite{Kawata2009}\cite{Parimi2003}. Deriving the exact resonances in terms of the properties of the material tremendously enhances the proper choice of its components by calculating its physical parameters based on the desired resonance effect. 

Functional analytic methods to approximate resonances' scatterers with high refractive indices are surveyed in~\cite{Ammari2021}, and quasi-resonances for single and multiple particles for which the corresponding eigenfunctions are uniformly constant are studied in~\cite{Challa2019}.

In this paper, we consider a three dimensional dielectric of Lipschitz boundary centered at the origin and placed in a vacuum. We developed formulas of resonances satisfying:  
$$\Delta u+k^2 (1+\eta)u=0$$

\noindent where $k$ is the wave number, $\eta$ is the medium's susceptibility coefficient, and $u$ is the scalar field. We denote $\lambda=k^2$ the spectral parameter associated with the integral non-linear eigenvalue problem defined in the next section in formula \eqref{eq:nonlinL}. In section \ref{asymsec}, we derive an asymptotic formula of resonances for high contrast nanoparticles of any shape, which is a development of the work in~\cite{Meklachi2018} when $\eta$ is constant. Section \ref{exact_for} presents exact resonances in a spherical ball, small and non-small, for radially symmetric fields in terms of its radius and dielectric susceptibility for any $\eta$  (not necessarily high). Towards the end of section \ref{exact_for}, a simplified version, also exact, is obtained for high contrast nanospheres. Section \ref{spec_sec} combines the previous results to shed light on the behavior of higher order correction terms in the full asymptotic expansion for high contrast nanospheres. 
\section{Formula of resonances of small volumes with high constant dielectric susceptibility}\label{asymsec}
Consider a 3D ball $B$ that contains the origin with arbitrary shape and dielectric susceptibility $\eta(x)$. We will derive an asymptotic formula for scattering resonances $\lambda_h$ on the scaled down volume $hB$ with a magnified dielectric susceptibility $$\eta=\chi_{hB}\frac{\eta_0}{h^2} .$$The field $u$ satisfies Helmholtz equation:
\begin{equation}
 \label{helm}
\Delta u+k^2(1+\eta)u=0
\end{equation}
where $k$ is the wave number and $\lambda=k^2$ is the spectral parameter.
The integral form of the solution of \eqref{helm} given by Lippmann-Schwinger is
\begin{equation}\label{lip}
u(x)=u_i(x)+k^2\int_{hB}\frac{\eta_0(y)}{h^2}G(x,y)u(y)dy
\end{equation} 
where $u_i$ is the incident field and $G$ is the Green's function in 3D defined as:
$$G(x,y)=\frac{1}{4\pi}\frac{\exp({i\sqrt{\lambda}|x-y|)}}{|x-y|}$$ 
The change of variables $x=h\tilde{x}$ and $y=h\tilde{y}$ for $x \in \mathbb{R}^3 $ transforms equation \eqref{lip} to the domain $B$ as follows:
\begin{equation}\label{lipman}
u(h\tilde{x})=\tilde{u}(\tilde{x})=u_i(h\tilde{x})+\frac{1}{4\pi}k^2\int_{B}\eta_0(h\tilde{y})\frac{\exp({i\sqrt{\lambda}h|\tilde{x}-\tilde{y}|)}}{|\tilde{x}-\tilde{y}|}\tilde{u}(\tilde{y})d\tilde{y}
\end{equation}
For small enough $h$ we have
$$\tilde{u}(\tilde{x})\approx u_i(0)+\frac{\eta_0(0)}{4\pi}k^2\int_{B}\frac{1}{|\tilde{x}-\tilde{y}|}\tilde{u}(\tilde{y})d\tilde{y}$$
Now let's define $u_0$ to solve
\begin{equation*}
u_0(\tilde{x})=u_i(0)+\frac{\eta_0(0)}{4\pi}k^2\int_{B}\frac{1}{|\tilde{x}-\tilde{y}|}u_0(\tilde{y})d\tilde{y}
\end{equation*}
or
\begin{equation*}
u_0=u_i(0)+k^2T_0u_0
\end{equation*}
where
\begin{equation}\label{eq:T00}
T_0(u)(\tilde{x})=\frac{1}{4\pi}\int_B  \eta_0(0)\frac{u(\tilde{y})}{|\tilde{x}-\tilde{y}|}d\tilde{y}.
\end{equation}
The resonances $\lambda_0$ are values for which \eqref{lip} has a nontrivial solution with no incident wave, hence they satisfy the eigenvalue problem   
\begin{equation}\label{eig1} 
\lambda_0 T_0(\lambda_0)u_0=u_0
\end{equation}
Similarly, rewriting \eqref{lipman}, the nonlinear eigenvalues $\lambda_h$ satisfy the eigenvalue problem 
\begin{equation}\label{eq:nonlinL}
\lambda_h T_h(\lambda_h)u_h=u_h
\end{equation}  
where
\begin{equation}\label{eq:Th1}
T_h(\lambda)(u)(\tilde{x})=\frac{1}{4\pi}\int_B  \eta_0(h\tilde{y})\frac{\exp({i\sqrt{\lambda}h|\tilde{x}-\tilde{y}|)}}{|\tilde{x}-\tilde{y}|}u(\tilde{y})d\tilde{y} \qquad
\end{equation}
 Equation \eqref{eig1} is a linear eigenvalue problem where $T_0$ is a self adjoint compact operator with positive eigenvalues. 
Computing $\lambda_0$ is a first step towards approximating $\lambda_h$. 
For practical purposes, assume $\eta_0$ is constant.
Theorem $2.1$ in reference [1] gives a first order asymptotic formula:
   \begin{equation*}
	\lambda_h=\lambda_0+{\lambda_0}^2\large\left\langle(T_0-T_h(\lambda_0))u_0,u_0\large\right\rangle+R_2(h)
\end{equation*} 
where $R_2(h)=\mathcal{O}(h^2)$.
\\ \\
First order asymptotic expansion of $\large\left\langle(T_0-T_h(\lambda_0))u_0,u_0\large\right\rangle$ yielded a general formula for approximating scattering resonances in high contrast small volumes with any arbitrary shape, which writes:
\begin{equation*}
	\lambda_h=\lambda_0-i{\lambda_0}^\frac{5}{2}\eta_0{U_0}^2h+R_1(h)+R_2(h) .
\end{equation*}  
where 
\begin{equation*}
	U_0=\int_B u_0(x)dx  \quad \text{and} \quad R_1(h)=\mathcal{O}(h^2).  
\end{equation*}
A full expansion formula of $\lambda_h$  requires recovering the terms $R_1$ and $R_2$ to any given order. 
In this paper, we are going to provide a full expansion of $R_1$ for high contrast small volumes of any geometric shape. However, the term $R_2$ is fully derived in this work for only the case of high contrast nano-spheres.
For constant $\eta_0$ we have
\begin{align*}
	(T_h(\lambda)-T_0)(u)=\frac{\eta_0}{4\pi}\int_B\big(\exp({i\sqrt{\lambda}h|x-y|)}-1\big)  \frac{u(y)}{|x-y|}dy
\end{align*}   
Taylor expansion on the function  $h\rightarrow \exp({i\sqrt{\lambda}h|x-y|)}$ to an order $n$ provides:
\begin{align*}
	(T_h(\lambda)-T_0)(u)&=\frac{\eta_0}{4\pi}\int_B\left(\sum_{k=1}^{k=n}\frac{\left(i\sqrt{\lambda} \left|x-y\right|\right)^k}{k!}h^k +o(h^n)\right)  \frac{u(y)}{|x-y|}dy\\
	&=\frac{\eta_0}{4\pi}
	\int_B\left(\sum_{k=1}^{k=n}\frac{\left(i\sqrt{\lambda}\right)^k}{k!} \left|x-y\right|^{k-1}h^k \right) u(y)dy+o(h^n)
\end{align*}   
\begin{align*}
&\large\left\langle(T_h(\lambda_0)-T_0)u_0,u_0\large\right\rangle=\\
&\frac{\eta_0}{4\pi}\sum_{k=1}^{k=n}{\lambda_0}^\frac{k}{2}\frac{(ih)^k}{k!}\int_{B}\int_{B}|x-y|^{k-1}{u_0}(x){u_0}(y)dxdy+o(h^n)\\
&=i\frac{\eta_0}{4\pi}{\lambda_0}^\frac{1}{2}{U_0}^2h+\frac{\eta_0}{4\pi}\sum_{k=2}^{k=n}{\lambda_0}^\frac{k}{2}\frac{(ih)^k}{k!}\int_{B}\int_{B}|x-y|^{k-1}{u_0}(x){u_0}(y)dxdy+o(h^n)\\
&=i\frac{\eta_0}{4\pi}{\lambda_0}^\frac{1}{2}{U_0}^2h-\frac{R_1(h)}{\lambda_{0}^2}
\end{align*}
A full expansion of $R_1(h)$ is then computed and is given in the following theorem:
\begin{thm:eigenvalueAppr1}\label{thm5}
Let $U$ be a domain bounded  away from the negative real axis in $\mathbb{C}$. Let $T_0$ and $T_h(\lambda)$  be two linear compact  operators  from $L^2(B)$ to $L^2(B)$ defined by \eqref{eq:T00} and \eqref{eq:Th1}, respectively. Let $\lambda_0\neq0$ in $U$ be a simple eigenvalue of $T_0$, and let $u_0$ be the normalized eigenfunction. Then for $h$ small enough and constant $\eta_0$, there exists a nonlinear eigenvalue $\lambda_h$ of $T_h$ satisfying the formula:
\begin{equation}\label{asymptotic}
	\lambda_h=\lambda_0-i\frac{\eta_0}{4\pi}{\lambda_0}^\frac{5}{2}{U_0}^2h+R_1(h)+R_2(h) .
\end{equation}  
where 
\begin{equation*}
	U_0=\int_B u_0(x)dx  \text{, } \quad R_2(h)=\mathcal{O}(h^2)
\end{equation*}
and 
\begin{equation}\label{R1}
	R_1(h)=-\frac{\eta_0}{4\pi}{\lambda_0}^2\sum_{k=2}^{k=n}{\lambda_0}^\frac{k}{2}\frac{(ih)^k}{k!}\int_{B}\int_{B}|x-y|^{k-1}{u_0}(x){u_0}(y)dxdy+o(h^n)
\end{equation}
\end{thm:eigenvalueAppr1}
\section{Exact resonances in a sphere for radially symetric fields}\label{exact_for}
Consider $B_h$ a ball of radius $r_h$ and coefficient of susceptibility $\eta_h$. In this section we derive the scattering resonances $\lambda_h$ of $B_h$ in a vacuum. We are using the notations $B_h$ and $r_h$ to be consistent with the application that follows the exact formula of $\lambda_h$. However, there is no assumption of small volume for $B_h$ nor a high contrast $\eta_h$.    
The governing equations with appropriate interface compatibility conditions satisfied by the modes are given by:
\begin{align}\label{helm2}
	&\Delta u_1+k_h^2(1+\eta_h)u_1=0 \mbox{  in } B_h\\
	&\Delta u_2+k_h^2u_2=0 \mbox{  in }\mathbb{R}^3\backslash \bar{B}_h \\
	& u_1=u_2 \mbox{  on } \partial B_h \label{cond1}\\
	&\frac{\partial u_1}{\partial\nu}=\frac{\partial u_2}{\partial\nu} \mbox{  on } \partial B_h \label{cond2}
\end{align}
For simplicity, we will restrict our attention to radially symmetric  modes which will allow us to exactly calculate resonances in terms of $h$.
In addition, the wave must be an outgoing wave at infinity and bounded inside the sphere. This gives the solutions:
\begin{align}
	& u_1=Aj_0\left(k_h\sqrt{1+\eta_h}r\right)\mbox{ in the dielectric}\label{bessel} \\
	& u_2=Bh_0(k_hr) \mbox{ in vacuum}\label{hankel}
\end{align}   
where $j_0$ and $h_0$ are the spherical Bessel of the first kind and the spherical Hankel function of the first kind, respectively, and $A$ and $B$ are constants.
Condition \eqref{cond1} gives:
\begin{equation*}
	Aj_0\left(k_h\sqrt{1+\eta_h}r_h\right)=Bh_0(k_h r_h)
\end{equation*}
Using \eqref{cond2}, we obtain the equation:
\begin{equation*}
	\sqrt{1+\eta_h}h_0(k_hr_h){j_0}'\left(k_h\sqrt{1+\eta_h}r_h\right)={h_0}'(k_hr_h)j_0\left(k_h\sqrt{1+\eta_h}r_h\right)
\end{equation*}
which can be solved for $k_h$ by symbolic MATLAB package:
\begin{equation*}
	k_h=-\frac{i}{r_h\sqrt{\eta_h+1}}\log \left(i\frac{ \sqrt{\eta_h+1}+1}{\sqrt{\eta_h}}\right)
\end{equation*}

The wave number is then extracted from the complex logarithmic function:
\begin{equation*}
k_h=\frac{1}{r_h\sqrt{\eta_h+1}}\left(\frac{\pi}{2}-i\log \left(\frac{ \sqrt{\eta_h+1}+1}{\sqrt{\eta_h}}\right)\right)
\end{equation*}
and therefore the scattering resonances are:
\begin{equation*}
	\lambda_h=k_h^2=-\frac{1}{r_h^2(\eta_h+1)}\log^2 \left(i\frac{ \sqrt{\eta_h+1}+1}{\sqrt{\eta_h}}\right)
\end{equation*}
and the resonance with the first positive real part is:
 \begin{equation*}
 \lambda_h=\frac{1}{r_h^2\left(\eta_h+1\right)}\left(\frac{\pi}{2}-i\log \left(\frac{ \sqrt{\eta_h+1}+1}{\sqrt{\eta_h}}\right)\right)^2
 \end{equation*}
The results are summarized in the following theorem:
\begin{thm:eigenvalueAppr2}\label{thm6}
Let $B_h$ be a spherical ball in a vacuum of radius $r_h$ with a coefficient of susceptibility $\eta_h$. For radially symmetric modes, the wave number is given by: 
\begin{equation*}
k_h=\frac{1}{r_h\sqrt{\eta_h+1}}\left(\frac{\pi}{2}-i\log \left(\frac{ \sqrt{\eta_h+1}+1}{\sqrt{\eta_h}}\right)\right)
\end{equation*}
 and the associated scattering dielectric resonance is given by:
 \begin{equation*}
	\lambda_h=\frac{1}{r_h^2\left(\eta_h+1\right)}\left(\frac{\pi}{2}-i\log \left(\frac{ \sqrt{\eta_h+1}+1}{\sqrt{\eta_h}}\right)\right)^2
\end{equation*}
\end{thm:eigenvalueAppr2}
Note that there was no assumption needed of small volume for $B_h$ nor of a high contrast index $\eta_h$.  
The following corollary is an immediate consequence for a nanosphere with high refractive index. 
\begin{cor:nano}\label{cor2}
Let $B_h$ be a nanospherical ball of radius $h$ and high coefficient of susceptibility  $\eta_h=\chi_{hB}\frac{\eta_0}{h^2}$. For radially symmetric modes, the wave number is given by the formula:
\begin{equation*}
k_h=\frac{1}{h\sqrt{\eta_h+1}}\left(\frac{\pi}{2}-i\log \left(\frac{ \sqrt{\eta_h+1}+1}{\sqrt{\eta_h}}\right)\right)
\end{equation*}
and the scattering dielectric resonances of $B_h$ in vacuum are given by the formula:

\begin{equation}\label{exact_cor}
	\lambda_h=\frac{1}{\eta_0+h^2}\left(\frac{\pi}{2}-i\log \left(\frac{ \sqrt{\eta_h+1}+1}{\sqrt{\eta_h}}\right)\right)^2
\end{equation}
\end{cor:nano}

For $\eta_0=1$, The wave number can be simplified to:
$$
k_h=\frac{\pi }{2 \sqrt{h^2+1}}-\frac{i \sin ^{-1}(h)}{\sqrt{h^2+1}}
$$
and the resonance as:
\begin{equation}\label{exact}
\lambda_h=\frac{\pi ^2}{4 \left(h^2+1\right)}-\frac{\sin ^{-1}(h)^2}{h^2+1}-\frac{i \pi  \sin ^{-1}(h)}{h^2+1}
\end{equation} 

\section{Spectral analysis of scattering resonances of high contrast nanospheres}\label{spec_sec}
Consider a sphere shaped scatterer $B_h$ centered at the origin of radius $r_h=hr_0$ and coefficient $\eta_h=\frac{\eta_0}{h^2}\chi_{B_h}$, where $r_0$ and $\eta_0$ are, respectively, the radius and susceptibility coefficient of the scaled ball $B$ that we will set to 1 $(r_0=\eta_0=1)$. In this section, we will present a scheme to compute $R_1$ and $R_2$ in the asymptotic formula for high contrast nanospheres given in theorem \eqref{thm5}. We will compute exactly the first two terms of $R_1(h)$,  accompanied with a general methodology to integrate $$\int_{B}\int_{B}|x-y|^n{u_0}(x){u_0}(y)dxdy$$ and evaluated explicitly for $n=1$ and $n=2$. The other terms, when $n>2$, can be computed similarly, although the algebra becomes messy and one might resort to symbolic toolboxes. The exact computation of $R_1(h)$ coefficients enables the exact derivation of  $R_2(h)$, as Formula \eqref{asymptotic} rewrites 
$$R_2(h)= \lambda_h-\lambda_0+i\frac{\eta_0}{4\pi}{\lambda_0}^\frac{5}{2}{U_0}^2h-R_1(h)$$ \label{R2}
where $\lambda_h$ is given by \eqref{exact_cor} or \eqref{exact} when $\eta_0=1$.

This analysis can be done to any order and, therefore, brings insight to the contribution of $R_1(h)$ and $R_2(h)$, components of the full expansion of $\lambda_h$ for high contrast nanospheres. 
First, we solve for $\lambda_0$ and $u_0$ in the limiting eigenvalue problem \eqref{eig1}:
\begin{equation*}
	\frac{\lambda_0}{4\pi}\int_B  \frac{u_0(y)}{|x-y|}dy=u_0(x).
\end{equation*}      
This equation was solved in~\cite{Meklachi2018} and we obtained:$$\lambda_0=\frac{\pi^2}{4}$$ and the normalized eigenfunction $u_0$ in the $L^2(B)$ norm is given by:
\[
u_0(r\hat{\theta})=\frac{1}{\sqrt{2\pi}}\frac{\sin{\frac{\pi }{2}r}}{r}=\frac{1}{\sqrt{2\pi}}\frac{\sin{\frac{\pi }{2}|x|}}{|x|}
\]\label{u0}
where $\hat{\theta}$ is the unit radial vector and $x=r\hat{\theta}$. 
We then have the first order correction term: $$\frac{\eta_0}{4\pi}{\lambda_0}^\frac{5}{2}{U_0}^2=\pi$$
Note that the first order asymptotic approximation 
$$\lambda_h=\left(\frac{\pi}{2}\right) ^2- i \pi h+\mathcal{O}(h^2)$$ given by theorem  \eqref{thm5} is an exact match with the first two terms in the Taylor expansion of $\lambda_h$ given by \eqref{exact}:
	\begin{equation}\label{exactLm}
		\lambda_h=
		\frac{\pi ^2}{4}-i \pi  h+\left(-1-\frac{\pi ^2}{4}\right) h^2+\frac{7}{6} i \pi  h^3+\left(\frac{4}{3}+\frac{\pi ^2}{4}\right) h^4+ ...
	\end{equation}
 
\subsection{Computation of integrals $\int_{B}\int_{B}|x-y|^n{u_0}(x){u_0}(y)dxdy$}
In this section, we will compute the coefficients
\[\int_{B} \int_{B} |x-y|^{n} u_{0}(x) u_{0}(y) dx dy \]
where $$u_0(x)=\frac{1}{\sqrt{2\pi}}\frac{\sin{\frac{\pi }{2}|x|}}{|x|}.$$
Let $\B{x}$ and $\B{y}$ be unit balls centered at the origin. Consider the spherical coordinates for $y\in \B{y}$: $\x{y} = |y|$, $\th{y}$ is the angle between $y$ and the $z$-axis, and $\ph{y}$ is the angle of rotation of $y$ around the $z$-axis. In the inner integral, for each fixed $y$, we use the coordinates for $x \in \B{x}$ given by $\x{x} = |x|$, $\th{x}$ is the angle between $x$ and $y$, and $\ph{x}$ is the angle of rotation of $x$ around the line through the origin and $y$. If $y$ is the origin, then we can just take the standard spherical coordinates on $\B{x}$.

So, we consider the inner integral first:
\begin{align*}
	&\int_{\B{x}} |x-y|^{\pow} u_{0}(x) dx \\
	& = \int_{0}^{1} \int_{0}^{\pi} \int_{0}^{2\pi} \left(\x{x}^2 + \x{y}^2 - 2\x{x}\x{y} \cos \th{x}\right)^{\frac{\pow}{2}} \frac{\sin \left(\frac{\pi}{2}\x{x}\right)}{\sqrt{2\pi} \x{x}} \cdot \x{x}^2 \sin \th{x} d \ph{x} d \th{x} d\x{x} \\
	& =  \sqrt{2\pi}\int_{0}^{1} \int_{0}^{\pi}  \left(\x{x}^2 + \x{y}^2 - 2\x{x}\x{y} \cos \th{x}\right)^{\frac{\pow}{2}} \sin \left(\frac{\pi}{2}\x{x}\right)  \x{x} \sin \th{x} d \th{x} d\x{x} 
\end{align*}
The inner integral of the last line becomes

\begin{align*}
	& \int_{0}^{\pi}  \left(\x{x}^2 + \x{y}^2 - 2\x{x}\x{y} \cos \th{x}\right)^{\frac{\pow}{2}} \sin \left(\frac{\pi}{2}\x{x}\right) \x{x} \sin \th{x} d \th{x} \\
	&  = \frac{1}{(\pow+2)\x{y}}\left(\x{x}^2 + \x{y}^2 - 2\x{x}\x{y} \cos \th{x})\right)^{\frac{\pow+2}{2}}  \sin \left(\frac{\pi}{2}\x{x}\right) \bigg|_{\th{x} = 0}^{\th{x} = \pi} \\
	& =  \frac{1}{(\pow + 2)\x{y}} \left[\left(\x{x}^2 + \x{y}^2 + 2\x{x}\x{y}\right)^{\frac{\pow+2}{2}} - \left(\x{x}^2 + \x{y}^2 - 2\x{x}\x{y} \right)^{\frac{\pow+2}{2}} \right]  \sin \left(\frac{\pi}{2}\x{x}\right)  \\
	&= \frac{1}{(\pow + 2)\x{y}} \left[\left(\x{x} + \x{y}\right)^{\pow+2} - \left|\x{x}- \x{y}\right|^{\pow+2} \right]  \sin \left(\frac{\pi}{2}\x{x}\right) \\
\end{align*}
This can be expanded for each $n$.  For example
\begin{align*}
	n=1 : & \quad\begin{cases} \frac{1}{3\x{y}} \left[6\x{x}^2 \x{y} + 2 \x{y}^3\right] \sin \left(\frac{\pi}{2}\x{x}\right) = \frac{1}{3}\left[6\x{x}^2 + 2 \x{y}^2\right] \sin \left(\frac{\pi}{2}\x{x}\right) & \x{x} > \x{y} \\
		\frac{1}{3\x{y}} \left[2 \x{x}^3+ 6\x{x} \x{y}^2\right] \sin \left(\frac{\pi}{2}\x{x}\right) & \x{x} \leq \x{y} \end{cases}
	\\
	n =2 : & \quad \frac{1}{4 \x{y}} \left[8 \x{x}^3 \x{y} + 8 \x{x} \x{y}^3 \right] \sin \left(\frac{\pi}{2}\x{x}\right) =  \left[2 \x{x}^3  + 2 \x{x} \x{y}^2 \right] \sin \left(\frac{\pi}{2}\x{x}\right)
\end{align*}
So, we integrate
\begin{align*}
	I & = \sqrt{2\pi} \int_{0}^{\x{y}} \frac{1}{(\pow + 2)\x{y}} \left[\left(\x{x} + \x{y}\right)^{\pow+2} - \left(\x{y}- \x{x}\right)^{\pow+2} \right]  \sin \left(\frac{\pi}{2}\x{x}\right) dx+ \\
	& \qquad
	\sqrt{2\pi} \int_{\x{y}}^{1} \frac{1}{(\pow + 2)\x{y}} \left[\left(\x{x} + \x{y}\right)^{\pow+2} - \left(\x{x}- \x{y}\right)^{\pow+2} \right]  \sin \left(\frac{\pi}{2}\x{x}\right) dx
\end{align*}
by parts.  This yields
\begin{align*}
	n = 1: & \dfrac{16\sqrt{2\pi}}{\pi^4 \x{y}} \left[ \pi^2 \x{y} - 4 \sin \left( \frac{\pi}{2} \x{y} \right) \right]\\
	n = 2: & \quad \dfrac{8\sqrt{2\pi}}{\pi^4}\left[\pi^2\x{y}^2+3\pi^2 - 24 \right]
\end{align*}

Now, integrating over $\B{y}$:
\begin{align*}
	\int_{\B{y}} I u_{0}(y) dy & = \int_{0}^{1} \int_{0}^{\pi} \int_{0}^{2\pi} I \frac{\sin \left( \frac{\pi}{2} \x{y}\right)}{\sqrt{2\pi} \x{y}} \cdot \x{y}^2 \sin \th{y} d \ph{y} d \th{y} d \x{y} \\
	& = 2\sqrt{2\pi} \int_{0}^{1} I \x{y} \sin \left(\frac{\pi}{2} \x{y}\right) d\x{y}
\end{align*}
which can be integrated by parts again. For instance, for $n=1, 2$
we obtain:
\begin{align}
	\int_{B} \int_{B} |x-y| u_{0}(x) u_{0}(y) dx dy&= \frac{128}{\pi^{3}} \label{int1}\quad \text{and}\\
	\int_{B} \int_{B} |x-y|^{2} u_{0}(x) u_{0}(y) dx dy &=\frac{768}{\pi^{5}} \left[\pi^2-8\right]\label{int2}
\end{align}
Integral values in \eqref{int1} and \eqref{int2} were verified by numerical integration using Monte Carlo method and returned a double precision. 
\subsection{computation of $R_1$ and $R_2$}
\begin{figure}[ht!]
	\centering
	\begin{subfigure}[b]{\textwidth}
		\includegraphics[width=1\linewidth]{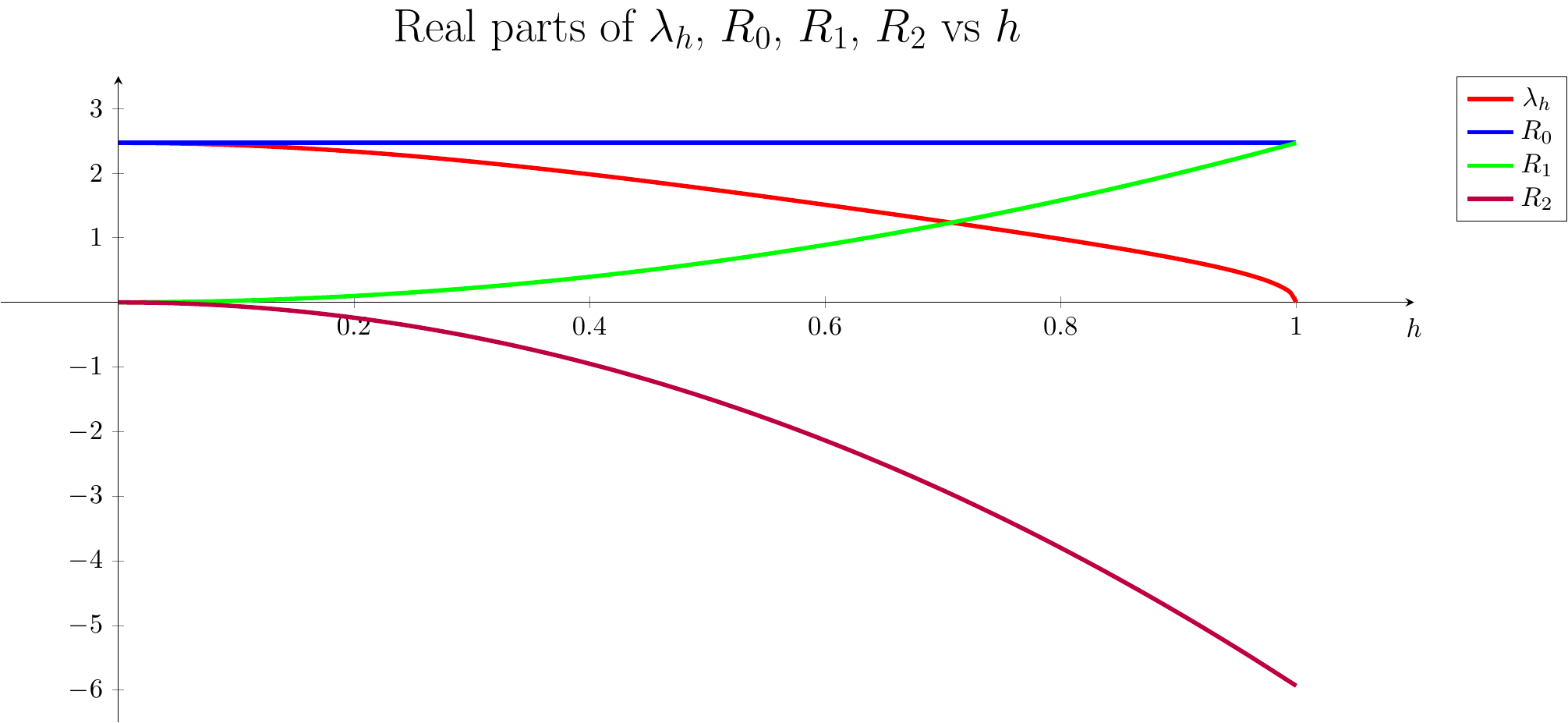}
		\caption{}
		\label{fig:Ng1} 
	\end{subfigure}
	
	\begin{subfigure}[b]{\textwidth}
		\includegraphics[width=1\linewidth]{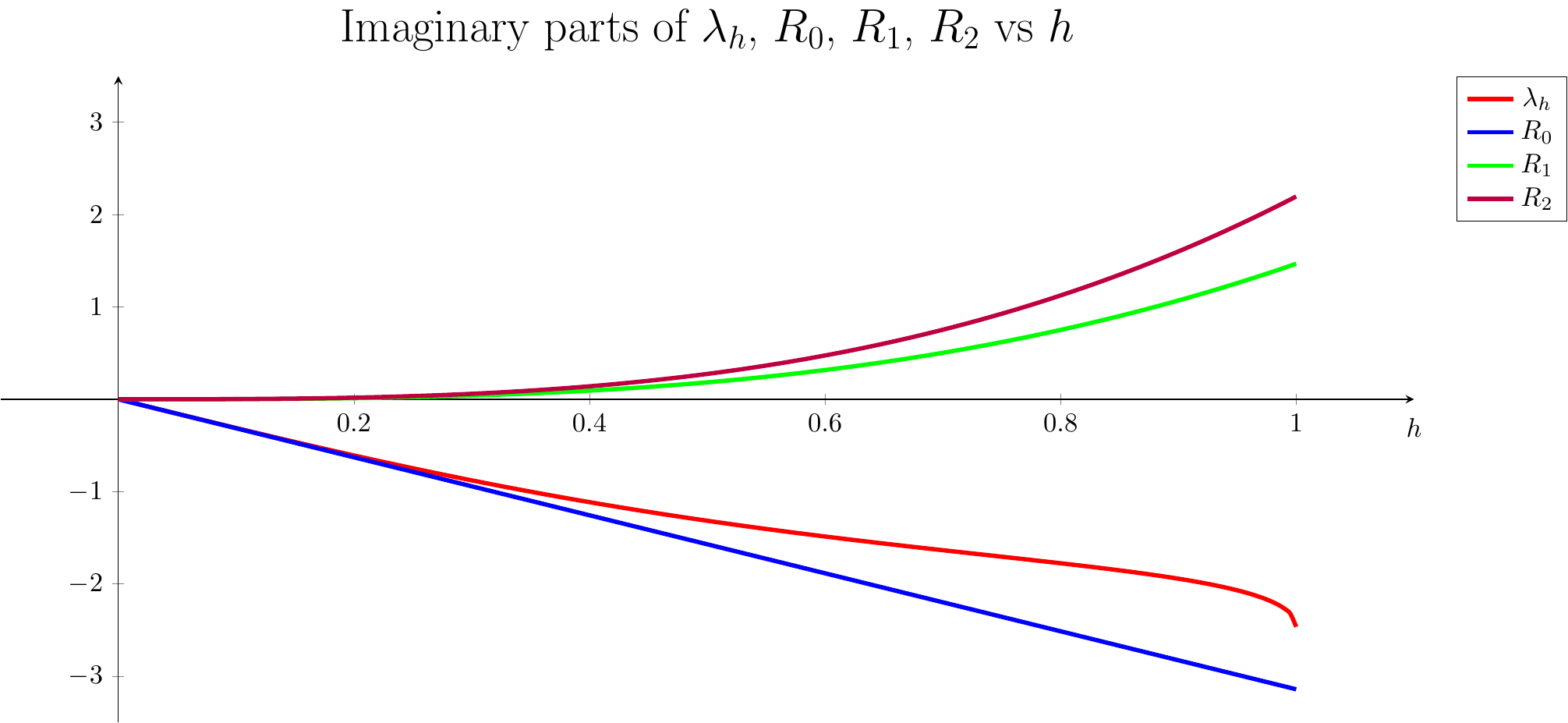}
		\caption{}
		\label{fig:Ng2}
	\end{subfigure}
	\caption{Exact $\lambda_h$ with respect to $h$ illustrated as a sum of the asymptotic corrections $R_0$, $R_1$ and $R_2$. The terms $R_1$ and $R_2$ are represented by the approximations in \eqref{R_1} and \eqref{R_2_2}.}
	\label{fig:test}
\end{figure}

Let \begin{equation}\label{R_1}
	R_1(h)=R_1^*h^2+R_1^{**}h^3+\mathcal{O}(h^4)
	\end{equation}

and
\begin{equation}\label{R_2_2}
 R_2(h)=R_2^*h^2+R_2^{**}h^3+\mathcal{O}(h^4)
\end{equation}\label{R_2}
Equation \eqref{R1} writes 
\begin{align*}
	R_1(h)&=
	-\frac{\eta_0}{4\pi}\lambda_0^3\frac{(ih)^2}{2}\int_{B}\int_{B}|x-y|{u_0}(x){u_0}(y)dxdy\\
	&-\frac{\eta_0}{4\pi}{\lambda_0}^{\frac{7}{2}}\frac{(ih)^3}{3!}\int_{B}\int_{B}|x-y|^2{u_0}(x){u_0}(y)dxdy+\mathcal{O}(h^4)
\end{align*}
Formulas \eqref{int1} and \eqref{int2} give:
\begin{align*}
R_1^*&=\frac{\pi^2}{4}\\
	R_1^{**}&=i\left(\frac{\pi^3}{4}-2\pi\right)
\end{align*}
To derive the first two terms of $R_2(h)$ we use expression \eqref{R1}	 along with the second and third order corrections of $\lambda_h$ in \eqref{exactLm}. We then obtain:
\begin{align*}
	R_2^*&=-1-\frac{\pi^2}{2}\\
	R_2^{**}&=i\left(\frac{19\pi}{6}-\frac{\pi^3}{4}\right)
\end{align*}
The asymptotic formula \eqref{asymptotic} rewrites $$\lambda_h=R_0(h)+R_1(h)+R_2(h)$$
where $$R_0(h)=\frac{\pi ^2}{4}-i \pi  h$$
To illustrate the contribution of $R_1(h)\approx R_1^*h^2+R_1^{**}h^3$ and $R_2(h) \approx R_2^*h^2+R_2^{**}h^3$ we simultaneously plot the real and imaginary parts of the exact $\lambda_h$ along with $R_0$ and the approximate values of $R_1$ and $R_2$, as shown in figure \eqref{fig:test}.

For nanospherical scatterers, plot \eqref{fig:Ng1} shows if $\lambda_h$ is unknown, and consequently so is $R_2$, then $R_1$ alone does not contribute to a better approximation of the real part of $\lambda_h$. Moreover, figure \eqref{fig:Ng2} shows that $R_1$ enhances the first order approximation of the imaginary part of $\lambda_h$ without the contribution of $R_2$. A generalization of the validity of latter observation on volumes with different geometries will be the subject of a future project. This analysis, therefore, suggests potential to expand the scope of the asymptotic formula in Theorem \eqref{thm5} to increase the order of approximation and to include non small dielectric scatterers of different shapes and of arbitrary susceptibility index that is not necessarily high.    
\newpage
\printbibliography

\end{document}